\documentclass{article}
\usepackage{graphicx}
\usepackage{authblk}
\usepackage{verbatim}
\usepackage{multirow}
\usepackage{array, tabularx}
\usepackage{subfigure}
\usepackage{comment}

\usepackage[a4paper, total={6.6in, 8in}]{geometry}
\usepackage{xr}
\usepackage[utf8]{inputenc}
\usepackage[T1]{fontenc}
\usepackage{amsthm}
\usepackage{amsmath}
\usepackage{amssymb}
\usepackage{siunitx}
\usepackage{bbding}
\usepackage{url}
\usepackage{booktabs} % For formal tables
\usepackage{xspace}
\usepackage{float}
\usepackage{multirow}
\usepackage{alltt}
\usepackage{color}
\usepackage{pifont}
\usepackage{makecell}
\usepackage{multicol}
\usepackage[super]{nth}

\let\origref\ref
\def\ref#1{\textnormal{\origref{#1}}}

\title{VoteLab: A Modular and Adaptive Experimentation Platform for Online Collective Decision Making}

\author[1]{Renato Kunz}
\author[2]{Fatemeh Banaie}
\author[2]{Abhinav Sharma}
\author[1]{Carina I. Hausladen}
\author[1]{Dirk Helbing}
\author[2]{Evangelos Pournaras}

\affil[1]{Computational Social Sciences, ETH Zurich, Switzerland\ \ \ \ \ \   \ \ \ \ \ \ \ \ \ \ \ \ \ \ \ \ \ \ \ \ \ \ \ \ \ \ \ \ \   \ \ \ E-mails:\{dirk.helbing,carina.hausladen\}@gess.ethz.ch, renato-kunz.msn.com}

\affil[2]{School of Computing, University of Leeds, Leeds, UK, \ \ \ \ \ \ \ \ \ \ \ \ \ \ \ \ \ \ \ \ \   \ \ \ \ \ \ \ \ \ \ \ \ \ \ \ \ \ \ \ \ \ \ \ \ \   \ \ \ \ E-mails: \{a.sharma1,E.Pournaras\}@leeds.ac.uk, F.Banaieheravan@hud.ac.uk}

\date{}

\begin{document}
	\maketitle
	\begin{abstract}
		\footnotetext[1]{$^*$ Corresponding author: Evangelos Pournaras, School of Computing, University of Leeds, Leeds, UK, E-mail: e.pournaras@leeds.ac.uk}

%Words: 235
Digital democracy and direct digital participation in policy making gain unprecedented momentum. This is particularly the case for preferential voting methods and decision-support systems designed to promote fairer, inclusive and legitimate collective decision-making processes for citizens' assemblies, participatory budgeting and elections. So far, a systematic human experimentation with different voting methods is cumbersome and costly. This paper introduces VoteLab, an open-source and well-documented platform for modular and adaptive design of voting experiments. It supports a visual and interactive building of reusable campaigns with different voting methods, while voters can easily respond to subscribed voting questions on a smartphone. A proof-of-concept with four voting methods and questions on COVID-19 have been used in an online lab experiment to study the consistency of voting outcomes. This demonstrates the Votelab capability to support rigorous experimentation of complex voting scenarios.

\end{abstract}

\section{Introduction}

Digital democracy initiatives with direct citizens' participation in decision and policy-making gain significant momentum, for instance, citizens' assemblies and participatory budgeting~\cite{Wellings2023,Helbing2023}. Limitations of current electoral systems as well as inaccurate or polarized voting outcomes of majority voting create the need to experiment with alternative preferential voting methods~\cite{Emerson2020}. This requires digital tools that are easy and trustworthy for voters, while the design of a campaign by researchers and policy-makers is simple, modular, and adaptive to different evaluation scenarios, offering flexibility to test a broad spectrum of voting methods. Although there are significant ongoing efforts in this direction \cite{Stanford,Consul,Decidim,Pournaras2020}, the existing voting and participation platforms have not materialized on all these capabilities. %materialized at this level. 

To change this, this paper introduces VoteLab, an open-source platform for modular and adaptive experimentation with different voting methods on smartphones. VoteLab allows users to visually and interactively design a voting campaign without writing a single line of code. Designers can even preview the users' voting experience in different smartphones before deployment. They can easily match voting questions to different voter groups, using assignments of tags/topics via a publish-subscribe system. VoteLab can collect useful meta-information to understand voting behavior such as voting time duration, time of choice, change of choices and feedback on voting outcomes. This allows one to conduct studies with between- and within-subjects designs, including factorial designs with different voting questions, different voting methods and different (treatment) groups. As a proof-of-concept, an online experiment is conducted to study four voting methods~\cite{voting_rules_carina_2024} in four voting questions related to the COVID-19 pandemic, i.e. in a polarized voting context. It is known from axiomatic results in social choice theory that voting outcomes may differ depending on the input method used~\cite{Arrow2012}. Accordingly, experimental insights are needed to understand better what are the factors that matter for voting outcomes, and which voting procedures are assessed by voters to be more favorable, trustworthy, and fair. Based on the collected data and experimental conduct, we conclude that VoteLab supports rigorous experimentation with complex collective decision-making scenarios. 
%% Replaced "very well" at the end with "efficiently" after VoteLab word, Removed efficiently
 
The main contributions of this paper are (i) a modular and adaptive modeling architecture for flexible experimentation with different voting methods; (ii) an open-source platform of VoteLab that implements the modeling architecture with a Web dashboard and an Android app; (iii) a proof-of-concept on COVID-19 to assess the practicality of VoteLab to support rigorous experimentation of complex voting scenarios; (iv) a software artifact demonstrator running on a virtual machine for reproducibility, assessment and engagement of the broader research community; (v) a comprehensive documentation of VoteLab for end-users and developers as well as a guide for the software artifact demonstrator~\cite{Artifact}. The rest of the paper is outlined as follows: Section 2 compares VoteLab with related work. Section 3 introduces the VoteLab architecture. Section 4 outlines the components of VoteLab. Section 5 illustrates the proof-of-concept of the software artifact. Section 6 concludes this paper and outlines future work.

\section{Comparison with related work}

Recent efforts focus on the implementation of participatory decision-making processes using digital voting platforms. Consul \cite{Consul} is such an open-source platform, developed by the Madrid city council for engaging public in decision-making processes such as making proposals or allocating public budget. Decidim~\cite{Decidim} is an open-source digital platform for citizen participation. These platforms make it possible to democratically organize campaigns for proposals, public meetings, decision-making discussions and also vote on the selected proposals. Stanford Participatory Budgeting (SPB)~\cite{Stanford} is used for budgeting problems rather than collaborative legislation and proposal submissions. While Web apps provide cross-platform compatibility and are easily accessible via Web browsers, their performance is not comparable to the ones of native apps~\cite{Char2011}. Moreover, these platforms support a limited number of voting methods and lack significant built-in functionality of meta-data collection for understanding voting behavior.

\begin{table*}[!htb]
\caption{Comparison of some popular digital participatory platforms for collective decision making.}
\label{tab:1} % Give a unique label
\centering
\resizebox{\textwidth}{!}{%
\begin{tabular}{ l c c c c c c c c}
\hline
  & SPB & Consul & Decidim & MVS & M-Vote & DApps & Smart Agora & VoteLab\\
  Criteria & \cite{Stanford}  & \cite{Consul}  & \cite{Decidim}  & \cite{Ign2017} & \cite{MobileV} & \cite{Evoting} & \cite{SmartAgora,Pournaras2020}& \\
\hline
Modular architecture& $\times$& \Checkmark &$\times$&$\times$&$\times$&$\times$& \Checkmark&\Checkmark\\
Adaptation& \Checkmark& $\times$& $\times$& $\times$& $\times$& $\times$& \Checkmark & \Checkmark\\
Simplicity& \Checkmark& $\times$& $\times$& $\times$& \Checkmark& \Checkmark& \Checkmark&\Checkmark\\
Metadata collection& \Checkmark& \Checkmark& $\times$& $\times$& $\times$& $\times$& \Checkmark & \Checkmark\\
User feedback& $\times$& \Checkmark& \Checkmark& $\times$& $\times$& $\times$& \Checkmark& \Checkmark\\
Thoroughly documented & $\times$ & $\times$ & $\times$ &  $\times$ & $\times$ &  $\times$ &  \Checkmark&\Checkmark \\
Number of voting methods & 5 & 2 & 1 & 1 & 1 & 1 & surveys & 7 \\
Native app & $\times$ & $\times$ & $\times$ & Android & mobile device&  mobile device & Android &Android \\
Verification method&\textit{Code \& SMS} & \textit{Census info \& SMS} & \textit{Code \& SMS} & \textit{SMS} & \textit{Fingerprint} & \textit{Phone number}& \textit{Code} &\textit{Email} \\
\hline
\end{tabular}
}
\end{table*}

%This performance gap arises from the additional layer of abstraction that web apps rely on to bridge the gap between different operating systems and devices.
% Moreover, these platforms only support a limited number of voting methods and they do not have significant built-in functionality for the collection of meta-data for the explanation and interpretation of the voting behavior.

Mobile Voting System (MVS)~\cite{Ign2017} is an open-source Android voting application. Registration and casting of votes is based on SMS messages. M-Vote~\cite{MobileV} is a mobile voting system utilizing fingerprint identification for enhanced security and authentication. DApps~\cite{Evoting} is also a digital voting system focusing on integrity, where identification is performed using voters' mobile phone numbers. Smart Agora~\cite{SmartAgora,Pournaras2020} is a crowd-sensing ubiquitous platform for outdoor living-lab experiments. It is designed for geolocated decision-making at points of interest, while providing capabilities for passive mobile sensor data collection. Complex crowd-sensing tasks are designed visually and interactively without write code. Smart Agora has also been studied in the context of verifying conditions for more informed decision-making on the blockchain. It is applied to Smart City domains such as cycling risk assessment~\cite{Pournaras2020}. 

There are several limitations in current digital decision-making approaches. The design of these platforms is complex with limited modularity (\textit{i.e., modular architecture}). They often require programming skills to obtain high-quality comprehensive data (\textit{i.e., simplicity, metadata collection}), with different implementations (\textit{i.e., native app}).
There is a lack of flexible platforms for voting experimentation, as existing tools are typically limited to specific voting scenarios (\textit{i.e., numbers of voting methods, verification method}). Meta-data about the voting choices such as recording the choice duration and evaluations of the voting results are often necessary to understand voting behavior (\textit{i.e. user feedback}). Moreover, platforms provide a varying flexibility in customizing and reusing voting questions and settings (\textit{i.e. adaptation}). Open-source digital voting platforms tend to be complex and inadequately documented (\textit{i.e., thoroughly documented}). Table 1 provides a comparative summary of some prominent platforms.

\section{System Architecture}

The system architecture of VoteLab is designed to create an adaptive system that facilitates the seamless integration of new voting mechanisms. 

\noindent \textbf{Comparison of Voting Methods.}
\noindent VoteLab provides an extensible testbed environment, enabling rigorous experiments to assess different voting mechanisms such as majority, approval, score or quadratic voting (currently 7 supported). By combining the collection of voting data and choice meta-data, VoteLab supports researchers and policy makers to study evidence-based decision making and design fairer, more expressive inclusive voting systems.

\noindent \textbf{Simplified Voting Experience.}
\noindent In VoteLab, voters can effortlessly create digital voting and data collection processes, using an intuitive visual interface. No coding is required, as the platform empowers users to visually design and implement complex workflows running on smartphones. For example, a community organization can use the platform to design a visually appealing ballot with clear instructions and options for voters. 

\noindent \textbf{ Tag Assignment System.}
\noindent Voters can automatically access voting questions and campaigns via a tag assignment system, representing categories of interest. This system is implemented as a publish-subscribe mechanism, which effectively determines who can perform which actions. For instance, it allows precise control over which city district can access specific voting questions. In this way, voting designers can create tailored campaigns for specific groups and communities.

\noindent \textbf{Multiple Voting Campaigns for Field Tests.}
\noindent The platform empowers researchers to create multiple reusable voting campaigns that involve repeated measurements and group comparisons (between and within subjects experimental designs). Researchers and practitioners can easily design, setup and run voting processes managed via a user-friendly graphical user interface.

\noindent \textbf{Behavioral Analysis and Decision-Making Insights.}
\noindent VoteLab supports the recording of initial choices made, their timing, changed decisions as well as the duration of decision-making processes. By collecting such metadata throughout the voting process, new insights can be gained to better understand, effectively design, and improve voting procedures. 

\noindent \textbf{Customizable and Seamless Feedback System.}
\noindent Votelab supports a built-in feedback system for gathering user opinions, ratings, or responses regarding voting outcomes and experiences. This feedback can assess voters' satisfaction and the legitimacy of the voting processes.

\noindent \textbf{Ubiquitous Online Voting}
Via VoteLab, voters engage in voting processes using personal devices they are already familiar with, without specialized or dedicated voting hardware. This promotes flexibility and convenience, enabling voters to cast their votes online anytime and from anywhere with Internet access.

\section{Key System Components}

The \textit{modular architecture} of VoteLab prioritizes the separation of system components and leverages API calls for the communication interface, enabling seamless integration, replacement, or addition of code components. This design approach ensures flexibility, adaptability, and scalability, allowing for the inclusion of new code segments, such as an iOS application, or the integration of additional voting methods without disrupting the existing framework architecture. 

\begin{figure}[!htb]
\centering
  \includegraphics[width=0.95\columnwidth]{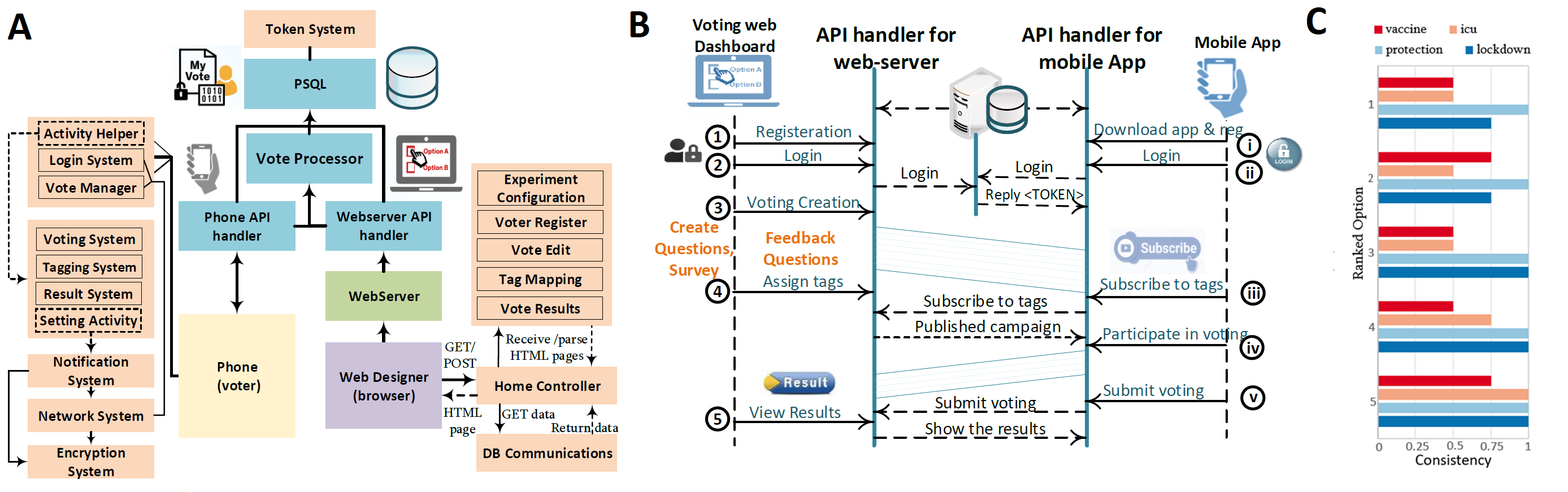}
  \caption{(A) VoteLab architecture. (B) VoteLab life cycle. (C) Consistency of voting outcomes.}
  \label{fig:merged}
\end{figure}

VoteLab is implemented using Android Studio, utilizing Java, and Microsoft .NET for the Web application. Figure~\ref{fig:merged}a depicts the architecture, which consists of three interactive parts: (i) an Android application enabling voters to actively participate in elections, (ii) a database server responsible for the storage and management of the collected data, (iii) a dedicated Web dashboard supporting the voting design with an intuitive and user-friendly interface. 

%In the rest of this section, we illustrate a detailed description of each component, emphasizing its distinct functionalities and features.

%For several questions, different input methods deliver different voting outcomes, but the consistency was 50\% or greater. The aggregation method (the way votes are summarized to determine the outcome) is also a significant part of the voting method that can be varied and may influence the outcome as well~\cite{Arrow2012,Benade2023}.

\noindent \textbf{Database Server and its Components}.
The central system component is the \textit{database server}, the hub for all communication and interactions. It plays a key role in handling changes, updates, and aggregations related to votes, i.e., the process of determining the overall voting outcome. The database server comprises three components: (i) a PostgreSQL database, (ii) API handlers for the phone application and Web server, and (iii) a vote processor. API handlers manage and translate external requests into PostgreSQL queries, ensuring effective communication with the database. Votes are processed by the vote processor, which generates the corresponding voting results. The processor regularly monitors the server for the closing date of voting, automatically calculates results, and stores them in the database upon completion of the voting period. If there is a need for calculating voting results other than the specified closing date, the API handlers request the calculation of results by establishing an open connection to the processor. 
%%This flexibility allows for on-demand result generation beyond the predefined closing date. 

\noindent \textbf{Voting Management Dashboard}.
The \textit{Web server component} encompasses the data and API calls within an intuitive Web interface, which can be used to design and deploy voting campaigns, as well as assign tags to voting questions, see Figure 2(a). Tag assignment involves associating specific tags or categories with voting experiments, allowing for easy categorization and organization of voting campaigns based on different criteria or themes. This enables efficient filtering and analysis of voting results using the assigned tags.

\begin{figure}[!htb]
    \centering
    % First Subfigure
    \subfigure[]{
        \includegraphics[height = 0.25\textwidth,width=0.4\textwidth]{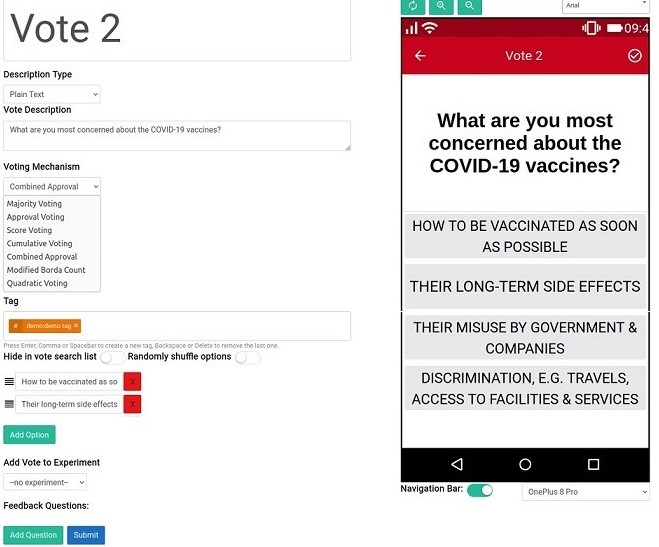}
    }
    \hfill
    % Second Subfigure
    \subfigure[]{
        \includegraphics[height=0.25\textwidth,width=0.18\textwidth]{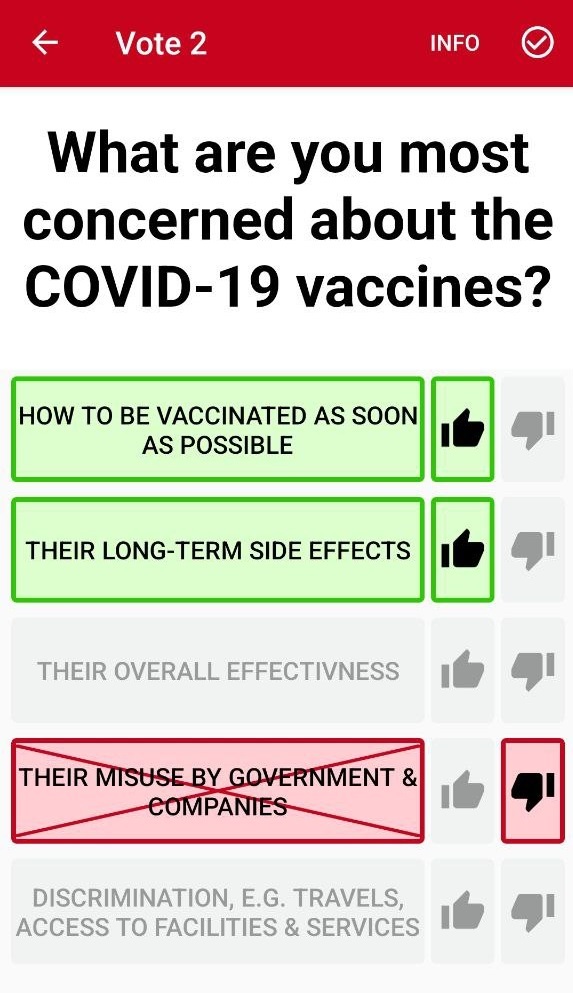}
    }
    \hfill
    % Third Subfigure
    \subfigure[]{
        \includegraphics[height=0.25\textwidth,width=0.2\textwidth]{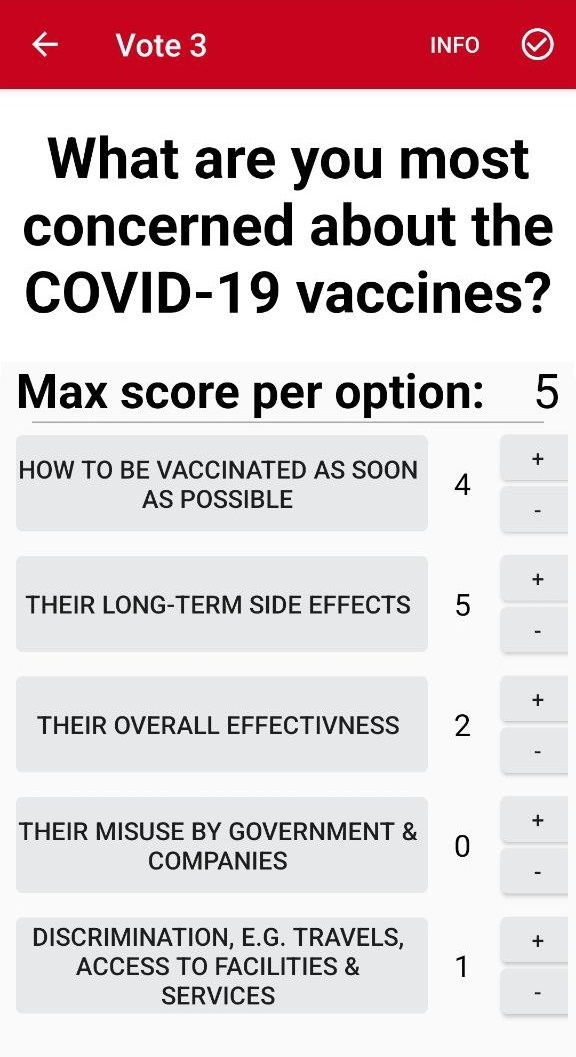}
    }
        \caption{VoteLab user interface: (a) Voting management dashboard. (b) and (c) show two example polls answered by the app user via two different voting methods: combined approval and score voting.}
    % \label{fig:threeFigures}
\end{figure}

\noindent \textbf{Voter Interface}. The VoteLab \textit{Android application} provides an intuitive interface for voters to actively participate in voting and experiments, see Figure 2, panels (b) and (c). Using the smartphone app, voters can access information about ongoing voting campaigns and experiments, view relevant details, and submit their responses. The platform uses the tag assignment system to match voters with voting questions. For example, if a voting campaign is related to a specific demographic or geographical region, voters assigned to the corresponding tags receive notifications and updates relevant to their specific group. This ensures that voters can receive tailored information and get opportunities to engage in voting processes that are more relevant to them. It also allows designers to easily create special group treatments to study voting behavior, presumably within the scope of proper ethics approvals. 

\noindent \textbf{VoteLab Workflow}. Figure~\ref{fig:merged}b illustrates the lifecycle of a voting experiment, starting from the ballot design to the campaign deployment and the calculation of the voting results. (1) The process begins with the ballot designers creating an account on the Web dashboard and (2) subsequently logging in, to create and run voting campaigns. (3) Via an intuitive interface, designers can effortlessly create and customize voting experiments. (4) Voters, on the other hand, can easily access the created voting campaign, express their preferences, and (5) view the results, once the voting period concludes. Voters can provide valuable feedback on voting results. The dashboard allows users to assign tags and reuse voting processes, enabling the deployment of multiple voting campaigns at different time points with different or the same voting participants. This feature simplifies experimentation with several waves of a panel study.

% \begin{figure}[!htb]
%   \includegraphics[width=0.5\columnwidth]{lifecycle.pdf}
%   \caption{Life cycle of the voting platform.}
% \end{figure} 

Voters have the option to (i) download the Android app and (ii) log in, using valid credentials. (iii) Once logged in, they can participate in voting by selecting the relevant voting tags they wish to contribute. (iv) As such voters are granted access to the corresponding voting questions. The tagging system allows voters to easily engage with projects that align with their interests. The privacy of votes and voters' anonymity is preserved in this process.

\section{COVID-19 Online Experiments}

To assess the modularity and adaptation capabilities of VoteLab, a proof-of-concept lab study is illustrated, based on an online experiment with human subjects. The experimental design involves a level complexity that is hard to manage with existing platforms: (i) four voting methods, (ii) four voting campaigns, (iii) two experimental conditions, (iv) a within-subjects design with repeated measurements, and (v) the collection of meta-data. The study is performed in 2021 and is related to COVID-19. It aims to understand how different preference elicitation methods influence voting outcomes in a polarized voting context. The in-depth analysis of the collected data is not subject of this paper and performed in earlier work~\cite{voting_rules_carina_2024}. Nevertheless, we outline some key findings of our proof-of-concept study using VoteLab. The online experiment was preregistered~\cite{voting_rules_carina_2024} and received approval by the ETH ethics commission. It involved 120 participants, who voted on different questions via different voting methods offered by VoteLab. 

%Each question offered participants five options to choose from.
%, modeled as $\mathcal{O}=\{o_1, o_2, o_3, o_4, o_5\}$. 
The questions to be voted upon were the following: 
\textbf{(1)} \emph{What are you most concerned about the COVID-19 vaccines?} [\emph{vaccine}] ($o_1$) How to be vaccinated as soon as possible. ($o_2$) Their long-term side-effects. ($o_3$) Their overall effectiveness. ($o_4$) Their misuse by governments \& companies. ($o_5$) Discrimination, e.g. travels, access to facilities \& services. 
\textbf{(2)} \emph{Among COVID-19 patients, which criterion should grant one access to an intensive care unit?} [\emph{icu}] ($o_1$) Being the youngest. ($o_2$) Being the oldest. ($o_3$) No denial of vaccination. ($o_4$) No violation of lockdown rules. ($o_5$) No health self-damage, e.g. smoking, drugs, alcohol. 
\textbf{(3)} \emph{Which is the most effective protection measure against a COVID-19 infection?} [\emph{protection}] ($o_1$) Wearing a mask. ($o_2$) Physical distancing. ($o_3$) Vaccination. ($o_4$) Regular hand washing. ($o_5$) Maintaining a healthy lifestyle. 
\textbf{(4)} \emph{Which is the most significant problem that the lockdown has caused?} [\emph{lockdown}] ($o_1$) Economic recession \& unemployment. ($o_2$) Government control \& suppression of freedom. ($o_3$) Social segregation \& increased inequality. ($o_4$) Mental distress. ($o_5$) Reduced physical health condition.

% \begin{figure}[h!]
%   \includegraphics[width=0.7\columnwidth]{cons.jpg}
%   \caption{Consistency of voting outcomes for each ranked option. For several questions, different input methods deliver different voting outcomes, but the consistency was 50\% or greater. Note that the aggregation method (the way votes are summarized to determine the outcome) is also a significant part of the voting method that can be varied and may influence the outcome as well~\cite{Arrow2012,Benade2023}.} 
%   %A consistency of 1.0 for the 1st ranked option means that all voting methods elect the same option as 1st ranked.}
% \end{figure}

Each participant answered each question with four different voting/input methods, which vary in terms of the degree of freedom and expressiveness: 
%the following four input methods: 
(i) \emph{majority voting} ($mv\!\!=\!\{0, 1\}$), (ii) \emph{combined approval voting} ($cav\!\!=\!\{0, 0.5, 1\}$), (iii) \emph{score voting} ($sv\!\!=\!\{0, 0.2, 0.4, 0.6, 0.8, 1\}$), and (iv) \emph{modified Borda count} ($mbc\!\!=\!\{0, 0.2, 0.4, 0.6, 0.8, 1\}$, if all options are selected, otherwise adjusted accordingly). Each input method scores the options in a different way. The scores refer to the numerical values assigned to a choice and represent the degree of preference. Majority voting is the least flexible method. Via combined approval voting, participants express disapproval or support. Score voting allows even more fine-grained expression of preferences, as participants assign a score to each option. The modified Borda count encourages participants to make compromises: the more options selected, the higher the assigned score to each option. 

%Social choice theory distinguishes between the \textit{voting process} (i.e. input method) and the \textit{evaluation process} (i.e. aggregation method). For the online experiment, only the voting process was varied. The same \textit{evaluation process} was used across all voting processes: The highest sum of scores calculates the winner for $\{o_1,...,o_5\}$. 
%The objective is to compare the outcomes of these questions using four distinct voting methods. 
%This comparative analysis provides valuable insights into how the various voting methods perform in terms of capturing participants' preferences and opinions on the given Covid-19-related questions. 

\begin{table}[H]
\caption{Percentages of voters who ranked a specific option as their first choice. Columns represent four questions with the four voting methods (rows): Majority Voting (mv), Combined Approval Voting (cav), Score Voting (sv), Modified Borda Count (mbc). Within each question, voters were able to choose from 5 different options $(o_1, ..., o_5)$. The option ranked first by the majority of voters is presented in \textbf{bold}. 
%Scores of options for four COVID-19 questions and voting methods. Bold values: 1st ranked option.
}
\label{tab:2}       % Give a unique label
\centering
%\begin{tabular}{c c c c c c}
\begin{footnotesize}
\begin{tabular}{c@{\hspace{30pt}}c@{\hspace{30pt}}c@{\hspace{30pt}}c@{\hspace{30pt}}c@{\hspace{30pt}}c}
\hline
 Voting & & & Questions & \\ \cline{2-6}
 Methods & &\textit{vaccine} & \textit{icu} & \textit{protection} & \textit{lockdown}  \\
\noalign{}\hline\noalign{}
\quad  &$o_1$& 14.7\% & 26.4\% & 4.1\% & 31.1\% \\
\quad  &$o_2$& \textbf{31.8\%} & 16.5\% & 15.8\% & 2.4\% \\
\quad mv &$o_3$& 11.6\% & 7.9\% & 3.9\% & 25.4\% \\
\quad  &$o_4$& 18.6\% & 22.7\% & \textbf{62.9\%} & \textbf{36.6\%} \\
\quad  &$o_5$& 23.3\% & \textbf{26.5\%} & 13.3\% & 3.8\% \\
\cline{2-6}
\quad  &$o_1$ & 11.2\% & 21.8\% & 20.4\% & 22.6\% \\
\quad  & $o_2$& 18.6\% & \textbf{22.3\%} & 21.7\% & 13.5\% \\
\quad cav &$o_3$ & 19.8\% & 14.8\% & 14.9\% & 21.2\% \\
\quad  & $o_4$& \textbf{25.2\%} & 20.2\% & \textbf{22.4\%} & \textbf{23.5\%} \\
\quad  & $o_5$& 25.1 & 20.8\% & 20.6\% & 19.2\% \\
\cline{2-6}
\quad  &$o_1$ & 15.2\% & \textbf{22.9\%} & 19.9\% & 24.5\% \\
\quad  & $o_2$& 19.3\% & 21.4\% & 21.4\% & 11.3\% \\
\quad sv &$o_3$ & 17.9\% & 14.3\% & 13.2\% & 20.8\%\\
\quad  &$o_4$ & \textbf{24.9\%} & 19.8\% & \textbf{25.1\%} & \textbf{25.1\%} \\
\quad  &$o_5$ & 22.7\% & 21.6\% & 20.4\% & 18.3\% \\
\cline{2-6}
\quad  &$o_1$ & 16.6\% & \textbf{22.4\%} & 17.2\% & \textbf{25.3\%} \\
\quad  &$o_2$ & 19.6\% & 21.7\% & 22.6\% & 13.2\% \\
\quad mbc & $o_3$& 19.1\% & 15.4\% & 11.8\% & 21.3\% \\
\quad  &$o_4$ & 22.2\% & 19.9\% & \textbf{27.6\%} & 24.6\% \\
\quad  &$o_5$ & \textbf{22.5\%} & 20.6\% & 20.8\% & 15.6\% \\
\hline
\end{tabular}
\end{footnotesize}
%\begin{minipage}{\linewidth}
%Columns represent four questions, to be voted upon by the participants via the following four different voting methods (rows): Majority Voting (mv), Combined Approval Voting (cav), Score Voting (sv), Modified Borda Count (mbc).
%Within each question, the voter could assess 5 different options $(o_1, ..., o_5)$. The option that was ranked first by the majority of voters is printed in \textbf{bold}. 
%\end{minipage}
\end{table}

Table 2 illustrates the aggregate scores of each option for each question and voting method. Figure~\ref{fig:merged}c illustrates the consistency of voting outcomes for each of the 1st, 2nd, ..., 5th ranked option, derived from Table 2. For instance, consistency of 1.0 for the 1st ranked option means that all voting methods determine the same option as ranked 1st. 

A consistency of 0.5 for the 2nd ranked option means 2 out of 4 voting methods determine the same option as 2nd ranked. The results reveal the following: (i) Voting methods seem to show higher consistency with disagreements rather than agreements. (ii) For the 1st ranked option, the highest consistency is observed for the \textit{protection} question. (iii) The \textit{vaccine} question has the lowest mean consistency among all five ranked options. The highest mean consistency is found for the \textit{protection} question. 

Inconsistency may reflect potential voters' disagreements with voting outcomes, i.e. results likely to be disputed.
For highly polarizing questions, voting methods are expected less consistent. Using a more sophisticated voting method allows voters to express their preferences in a more differentiated way~\cite{voting_rules_carina_2024}. The extra information VoteLab collects (e.g. on decision times) can quantify the user's experience of the voting process. For instance, voter's decision time can be used as a proxy for choice complexity; changes in choices can indicate a weak preference.
%Such consistency may be considered to be a measure of robustness of the voting outcome with regard to the variation of the voting/input method. 

%The consistency of the voting methods can be compared across questions. 
% Accordingly, for highly polarizing questions, voting methods are expected to be less consistent. While choosing a simple, time-efficient method appears to be well suited for non-polarizing questions, for polarizing questions one should use a more sophisticated voting method, which allow voters to express their preferences in a more differentiated way~\cite{Carina2023}. The additional information VoteLab is able to collect (e.g. on decision times) can be used to quantify the user's experience of the voting process.

%For instance, the time that a voter took to decide can be used as a proxy for choice complexity; changes in choices can indicate a weak preference.
%information overload and choice complexity can be measured in VoteLab via the voting time, changes in voting choices or user feedback information

\section{Conclusions and Future Work}

This paper shows that VoteLab is a versatile open-source tool for adaptive experimentation with different voting and collective decision-making processes. Our results were obtained via a within- subjects design study on COVID-19 that involves four voting methods and questions as well as a collection of meta-data helping to explain and interpret the choices of voters. While there are other established tools for digital participation, VoteLab provides more voting methods to experiment with and a familiar, smartphone-based experience for users. 
%higher control level of the user experience in smartphones. 

However, VoteLab also has some limitations and offers opportunities for future work. The development of an iOS app and the option to vote directly from a browser is currently ongoing work. The same applies to experiments with different aggregation methods and applications to participatory budgeting. Other interesting study scenarios include policy-making for Smart Cities and sustainability applications, for instance, crowd-sensing of air/water quality~\cite{Griego2017,Mahajan2022}. The integration of VoteLab with Smart Agora~\cite{Pournaras2020,SmartAgora} will further provide new opportunities for online ubiquitous geolocated voting. Issues of mis- or disinformation and deception potentially influencing voting outcomes should also be addressed in the future. Another ambition of VoteLab is to include secure (blockchain-based) incentive and reward mechanisms~\cite{Dapp2021}.

\bibliographystyle{unsrt}
\bibliography{references}

\end{document}